\documentclass[10pt,aps,prl,superscriptaddress,twocolumn,showpacs]{revtex4-2}
\usepackage[version=3]{mhchem} 
\usepackage{graphicx}
\usepackage{dcolumn}
\usepackage{bm}
\usepackage{color}
\usepackage[colorlinks=true, linkcolor=blue, urlcolor=blue, citecolor=blue]{hyperref}

\begin{document}
\preprint{APS/123-QED}
\title{Formation and protection of an Eu-Ir surface compound below hexagonal boron nitride}

\author{Alaa Mohammed Idris Bakhit}
 \email{alaa.mohammed$@$ehu.eus}
 \affiliation{%
 Centro de F\'{\i}sica de Materiales (CSIC-UPV-EHU) and Materials Physics Center (MPC), 20018 San Sebasti\'an, Spain
}%

\affiliation{%
 Departamento de F\'{\i}sica Aplicada I, Universidad del Pa\'{\i}s Vasco UPV/EHU, 20018 San Sebasti\'an, Spain
}%
 \author{Khadiza Ali}
 \affiliation{%
Chalmers University of Technology, Göteborg, Chalmersplatsen 4, 412 96 Göteborg, Sweden 
}%
\affiliation{%
Department of Physics, BITS Pilani, Hyderabad Campus, Hyderabad, 500078, Telangana, India
}%

 \author{Frederik Schiller}
 \email{frederikmichael.schiller$@$ehu.es}
  \affiliation{%
 Centro de F\'{\i}sica de Materiales (CSIC-UPV-EHU) and Materials Physics Center (MPC), 20018 San Sebasti\'an, Spain
}%
 \affiliation{%
Donostia International Physics Center, 20018 Donostia-San Sebasti\'an, Spain
}%

\date{\today}

\begin{abstract}
Europium (Eu) intercalation below hexagonal boron nitride (hBN) on an Ir(111) substrate at various Eu coverages is investigated.
The structural and electronic properties were examined using low energy electron diffraction (LEED), scanning tunnelling microscopy (STM), x-ray photoelectron spectroscopy (XPS) and angle-resolved photoemission spectroscopy (ARPES). Depending on the deposition temperature, different superstructures, (5 $\times$ $M$), (5 $\times$ 2), and ($ \sqrt{3}$ $\times$ $\sqrt{3})R30^{\circ}$ with respect to the Ir substrate were identified by LEED. The (5 $\times$ $M$) superstructure ($M$ $>$ 2), at 0.10 monolayer (ML), preserved the hBN/Ir Moir{\'e} pattern and exhibited a unidirectional ordering of Eu atoms. At higher coverage of 0.26 ML, a (5 $\times$ 2) superstructure emerged, where excess Eu atoms diffused into the bulk and were analyzed as Eu in a tri-valent state. At the highest preparation temperature with a one-third ML Eu, the formation of a ($\sqrt{3}$ $\times$ $\sqrt{3})R30^{\circ}$ superstructure indicates the presence of a EuIr$_{2}$ surface alloy beneath the hBN layer, with di-valent Eu atoms suggesting potential ferromagnetic properties. Air exposure was used to evaluate the protection of the hBN layer, and the results indicate that the EuIr$_{2}$ surface alloy was partially protected. However, the hBN layer remained intact by intercalation and air exposure, as confirmed by ARPES analysis.
\end{abstract}
\maketitle

\section{\label{sec:1}Introduction}
2D materials grown on metal substrates~\cite{preobrajenski2007monolayer,auwarter2019hexagonal} have attracted significant attention due to their unique electronic properties and potential applications in nanoelectronics and spintronics. Among these, the intercalation of metals~\cite{demiroglu2019alkali} beneath these materials has emerged as a promising approach to tuning the electronic and magnetic properties of these systems~\cite{pustilnik2015solitons}. Additionally, 2D materials may serve as a protective layer, shielding underlying materials from ambient conditions. For instance, graphene~\cite{coraux2012air,martin2015protecting,weatherup2015long,cattelan2015nature, naganuma2020perpendicular,sutter2010chemistry,sokolov20202d,anderson2017intercalated} and hBN were used as protection layers \cite{liu2014quasi,caneva2017growth,jiang2017high,tang2021direct,holler2019air,zihlmann2016role,ma2022epitaxial}.
Especially rare-earth metals are known to be very reactive and oxidize easily under ambient conditions.
Europium belongs to this group of metals. In a solid compound, Eu may exhibit di-valent, tri-valent or even a mixed-valent state~\cite{Lawrence1981_RepProgPhys,Hossain2004_PRB,Nemkovski2016_PRB}. The di-valent state is known for its ferromagnetic properties, making Eu promising for spintronic applications. However, Eu is highly reactive in air, limiting its practical use. In such conditions, Eu will oxidize to the tri-valent Eu$_2$O$_3$, which does not reveal any more ferromagnetism. The intercalation of Eu has been investigated mainly for graphite and graphene~\cite{schumacher2014europium,schroder2016core,anderson2017intercalated,sokolov20202d,sokolov2021_JAlloyComp} but also recently below a hBN layer~\cite{bakhit2023ferromagnetic}. Nevertheless, not much attention was paid to the stability of Eu in such intercalation systems.

The intercalation of Eu under graphene/Ir(111) or hBN/Pt(111) systems gave rise to europium atoms with ferromagnetic behavior that depend on the amount of the intercalated Eu~\cite{schumacher2014europium,bakhit2023ferromagnetic}. In the former system, a case of a (2 $\times$ 2) superstructure was observed where the Eu-Eu distance doubles the Ir interatomic distance of the (111) face, amounting approx. 5.4 \AA. There, Eu atoms have been observed in a paramagnetic state~\cite{schumacher2014europium}. When the Eu atoms arrange in a ($\sqrt{3}$ $\times$ $\sqrt{3}$) superstructure corresponding to approx. 4.7 \AA~Eu-Eu distance, ferromagnetic behavior on Ir and Pt is obtained~\cite{schumacher2014europium,bakhit2023ferromagnetic}. The origin of the different magnetic properties was assigned mainly to the RKKY-type interaction between the Eu atoms mediated by valence band electrons. On one hand side, the distance between Eu atoms is one of the crucial parameters for the magnetic interaction; on the other hand, the influence of the valence bands from the Ir or Pt atoms is important~\cite{schumacher2014europium}. For the Eu-Ir system, one has to mention also the physical properties of bulk Eu-Ir compounds. Specifically, EuIr$_2$ drew attention in the late sixties of the last century as an initially described ferromagnetic compound~\cite{Bozorth1959_PR}, nevertheless, the magnetic susceptibility did not correspond to Eu$^{3+}$, the valence state of its constitutes~\cite{Bozorth1960_PR}. Only one year later, the discovery of EuO as a ferromagnetic compound resolved the puzzle, being the possible contamination of the latter the responsible for the measured ferromagnetic behavior in EuIr$_2$~\cite{Matthias1961_PRL, VanVleck1978_JLessComMet}. Interestingly, later EuIr$_2$ compound was found superconducting~\cite{Matthias1979_PLA}, confirming europium atoms in a tri-valent state. Furthermore, it should be mentioned that Eu in other bulk materials can exist with intermediate valence~\cite{Stockert2020_PRB}, either due to different crystallographic Eu sites as in Eu$_3$O$_4$~\cite{Rau1966_ActaCrys}, due to valent fluctuating system as in EuIr$_2$Si$_2$~\cite{Schulz2019_Npg} or due to non-integer Eu valence in intermediate valent systems that originate from hybridization between the 4f moments and the conduction electrons as in EuCu$_2$(Si$_x$Ge$_{1-x}$)$_2$~\cite{Hossain2004_PRB}. A last aspect to mention is that on the surface, however, nearly all Eu or Sm compounds are di-valent~\cite{Johansson1979_PRB,Schneider1983_RPB28}.

\begin{figure}
    \centering
    \includegraphics[width=8.6cm]{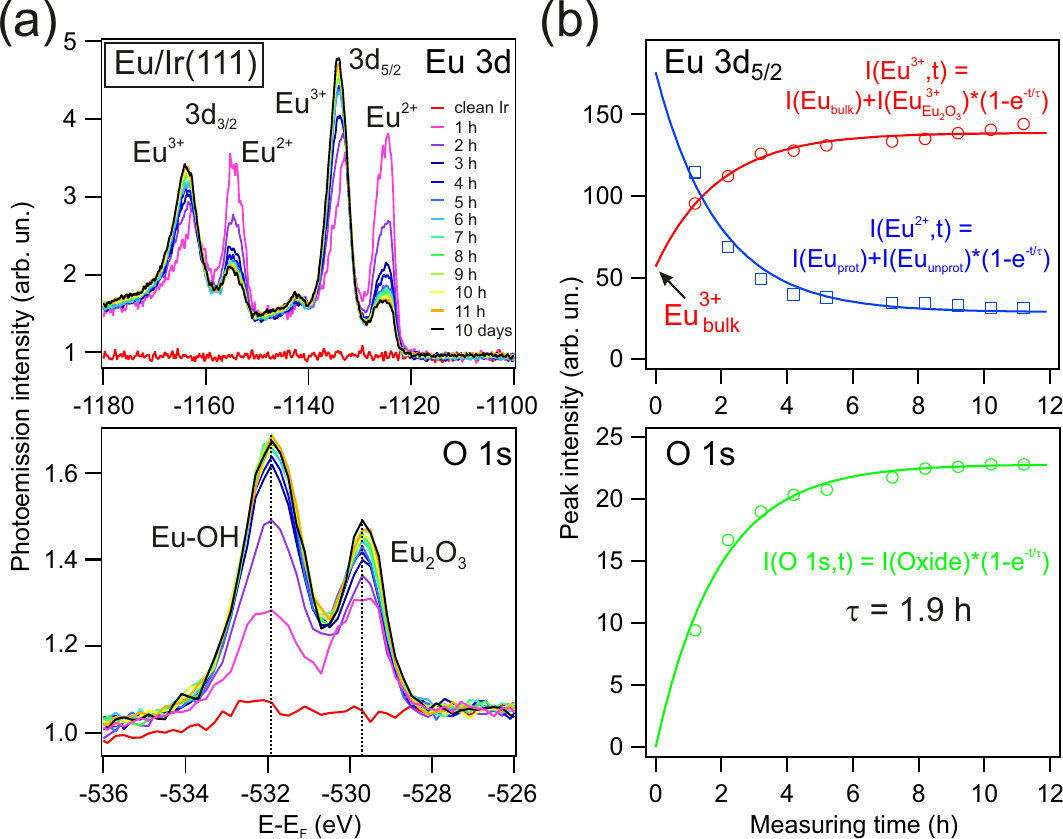}
    \caption{XPS analysis of Eu deposited onto Ir(111) crystal. (a) Eu 3d and O 1s X-ray photoemission spectra were obtained using Al K$_\alpha$ radiation with the sample kept in Ultra-High Vacuum (UHV) conditions. Oxidation takes place with the rest gas atoms in UHV. (b) Analysis of the peak intensities from the fitted spectra of (a). The solid lines are exponential fits of the peak areas as indicated in the equations. The lifetime $\tau$ resulted from the fit of the O 1s intensities (Eu$_2$O$_3$ and hydroxide) and was then applied to the Eu$^{2+}$ and Eu$^{3+}$ peaks.}
    \label{fig:1}
\end{figure}

As already mentioned and verified in the example of EuO contamination on EuIr$_2$, the protection of rare-earth metals in their metallic form is a major issue. In order to address the protection problem related to the very reactive rare-earth metals, first, the chemical properties of Eu deposited on an Ir(111) substrate in ultra-high vacuum conditions were analyzed by X-ray photoemission spectroscopy. For this, photoemission spectra of the Eu 3d and O 1s core level spectra were taken as a function of time, as shown in Fig.~\ref{fig:1}. The base pressure of the vacuum chamber during this experiment was $p$ = 3 $\cdot$ 10$^{-10}$mbar. The substrate was held at 630 K during deposition to allow for surface alloying. The XPS results reveal that following the deposition of approx. half a monolayer (ML) of Eu onto the Ir(111) substrate, Eu presented a mixture of valence states, namely di-valent and tri-valent Eu with binding energies of the leading 3d$_{5/2}$ peaks at $E-E_F$ = -1125.6 eV and -1134.7 eV, respectively. Here, one monolayer is defined as a complete closed Eu film on Ir. The presence of di-valent Eu at the surface, either due to an Ir-Eu surface alloy or due to individual Eu atoms at the surface, is detected. The tri-valent Eu emission may arise either from oxidized Eu$_2$O$_3$ or from Eu atoms that diffused into the bulk. Careful inspection of the peak form and energy reveals a change in that Eu$^{3+}$ emission, pointing to the fact that at start, Eu with a different environment is observed. The corresponding O 1s spectra exhibit two main components: Eu-OH at higher and Eu$_{2}$O$_{3}$ at lower binding energies~\cite{baltrus2019rare}. Additionally, a minor component corresponding to the adsorbed water was determined from the fit shown in supplemental material  \cite{SM} in Fig. 3. 
One observes that tri-valent Eu and O 1s increase with time while the di-valent Eu contribution diminishes.
One can fit the time evolution of the total O 1s intensity with an exponential curve by assuming that there is no oxide formed at deposition ($t$ = 0). This means that the Eu atoms at the surface oxidizes with the rest vacuum by
\begin{equation}
I(t)=I_{tot}\cdot\left ( 1-e^{-t/\tau}\right ).
\end{equation}
The only fitting parameter, $\tau$, is the mean lifetime of the di-valent Eu atoms at the surface and resulted in $\tau = (1.9\pm0.5)$ hours. Within this approx. 2 h time frame and taking into account the base pressure, one would accumulate 1.5 L of rest gas dosage on the sample and due to the high reactivity of Eu, oxidation and formation of the tri-valent Eu takes place. These findings highlight the rapid oxidation process of Eu that occurs without protective measures. To keep Eu in a metallic state, effective protection would be required.
Nonetheless, utilizing 2D materials such as hBN has proven effective in protecting Eu, as evidenced in the study of the formation of Eu-Pt ferromagnetic compounds under the hBN layer on Pt substrate~\cite{bakhit2023ferromagnetic}. Understanding how Eu behaves when intercalated beneath hBN is crucial for maximizing its potential in practical applications.

In this context, our study investigates the intercalation of Eu atoms beneath a monolayer of hBN on an Ir(111) substrate. By systematically varying Eu coverage and deposition conditions of Eu, we aim to uncover the structural and electronic transformations that occur.
Employing characterization techniques such as Low Energy Electron Diffraction (LEED), Scanning Tunneling Microscopy (STM), X-ray Photoelectron Spectroscopy (XPS), and Angle-Resolved Photoemission Spectroscopy (ARPES), we provide a comprehensive analysis of the Eu intercalation effects. Our findings reveal the formation of distinct superstructures, including (5 $\times$ $M$), (5 $\times$ 2), and $(\sqrt{3}$ $\times$ $\sqrt{3})R30^{\circ}$ depending on the Eu coverage and deposition temperature ($M$ $>$ 2). At low coverage (0.1 monolayer, ML), the (5 $\times$ $M$) superstructure preserves the hBN/Ir Moir{\'e} pattern with the unidirectional ordering of Eu atoms. Increasing the coverage leads to a (5 $\times$ 2) superstructure, with excess Eu atoms diffusing into the bulk as tri-valent Eu. At surface saturation coverage, the formation of a $(\sqrt{3}$ $\times$ $\sqrt{3})R30^{\circ}$ superstructure indicates the presence of a EuIr$_{2}$ surface alloy beneath the hBN layer, where di-valent Eu suggests potential ferromagnetic properties. Moreover, we assess the protective role of the hBN layer against air exposure, finding that the EuIr$_{2}$ surface alloy is partially protected. This investigation advances our understanding of Eu intercalation beneath hBN on Ir(111) and highlights the potential for developing new 2D ferromagnetic materials with tailored electronic and magnetic properties.

\section{\label{sec:2}Experimental}

Ir(111) single crystal (Mateck GmbH) was cleaned using multiple cycles of Argon ion sputtering (1.3 kV) at a pressure of 3 $\cdot$ $10^{-6}$ mbar at room temperature, followed by an annealing to 950 K. Occasionally, the sample was heated in an oxygen atmosphere at 5 $\cdot$ $10^{-8}$ mbar in order to remove carbon impurities, maintaining a temperature of 1120 K for 5 minutes, then flash-annealing to 1200 K for 4 minutes. Finally, a sharp hexagonal LEED pattern emerges, shown in Fig. \ref{fig:2}(a), demonstrating the three-fold symmetry characteristic of a pure fcc Ir(111) surface.
The synthesis of hBN on Ir(111) was achieved using the Chemical Vapor Deposition (CVD) technique. For this purpose, a clean Ir(111) substrate was heated and exposed to borazine precursor (B$_{3}$H$_{6}$N$_{3}$, KATCHEM spol. s r.o.) for 12 minutes at a pressure of 5 $\cdot$ $10^{-7}$ mbar. The substrate was maintained at temperatures between 1120 K and 1220 K, resulting in the growth of a uniform monolayer of hBN across the entire substrate. The formation of a Moir{\'e} pattern was observed, as illustrated in Fig. \ref{fig:2}(b).
Eu intercalation beneath the hBN on Ir(111) was carried out at three different temperatures: 800 K, 920 K, and 940 K, with corresponding LEED patterns shown in Figs.~\ref{fig:2}(c)-(e), respectively. Subsequent experiments were conducted to investigate oxidation protection. This was achieved through two methods: first, by directly exposing the sample to 1000 L of oxygen for 5 minutes; and second, by subjecting the sample to ambient pressure conditions (room temperature, 80$\%$ humidity) for 5 minutes. In both cases, followed by annealing to $T_\textnormal{Sample}$=745 K and $T_\textnormal{Sample}$=506 K, respectively.

The Eu deposition was carried out at a rate of approx. 0.2~\AA~ as measured previously by quartz microbalance at the deposition position. The evaporation of Eu on the sample was carried out for 15 min with the sample held at the indicated temperatures above. The pressure during evaporation was 3 $\cdot$ $10^{-9}$ mbar. Now, let us explain the method we used to estimate Eu coverage on the sample surfaces. Here, the focus was placed exclusively on di-valent Eu (Eu$^{2+}$), as this valence state is responsible for forming an ordered surface structure identifiable through LEED. In particular, the reference point for coverage was a well-established $(\sqrt{3} \times \sqrt{3})R30^{\circ}$ superstructure, which corresponds to a Eu coverage of one-third a monolayer (ML). This superstructure is considered the saturation limit for Eu on the surface. When Eu is deposited beyond this amount, the excess atoms either diffuse into the bulk of the Ir substrate or remain atop the hBN layer. In both scenarios, these additional Eu atoms typically exist in the tri-valent state (Eu$^{3+}$), which does not contribute to the ordered LEED pattern. To determine the Eu coverage in different preparations, we analyzed XPS spectra, specifically focusing on the intensity of the Eu$^{2+}$ 3d core level peak. This intensity was compared to that of the reference sample with one-third ML Eu, allowing us to calculate relative coverages. Importantly, only the Eu$^{2+}$ component was used in this calculation, as the Eu$^{3+}$ signal is associated with disordered or non-intercalated species. The coverage estimations obtained from XPS were validated by LEED and STM analyses, which provided structural confirmation of the surface arrangements. This combined approach allowed for a consistent and accurate quantification of Eu coverage across various preparations. 

All experiments were conducted under ultra-high vacuum (UHV) conditions at the Material Physics Center (MPC) in San Sebastian, Spain, maintaining a base pressure of 5 $\cdot$ $10^{-10}$ mbar or lower. The characterization of both structural and electronic properties encompassed a comprehensive array of techniques, including LEED, STM, ARPES, and XPS. LEED measurements utilized the conventional three-grid design from Omicron equipment. STM measurements were conducted in constant current mode using an Omicron VT-STM microscope, with image analysis processed via the WSXM program \cite{horcas2007wsxm}. Both techniques were performed at room temperature. Additionally, XPS spectra were obtained using a Specs Al K$_\alpha$ $\mu$-FOCUS 600 monochromator at $h\nu$ = 1486.6 eV.
ARPES measurements were carried out using a Specs UVS-300 discharge lamp with Specs TMM 304 monochromator utilizing He II$\alpha$ light at a photon energy of 40.8 eV. Electron detection was performed with a SPECS Phoibos 150 analyzer, configured for 200 meV energy and 2$^{\circ}$ angular resolution. Spectroscopic experiments were again performed at room temperature.

\section{\label{sec:3} E\lowercase{u}/\lowercase{h}BN/I\lowercase{r}(111) interface}
\begin{figure*}
    \centering
    \includegraphics[scale=0.46]{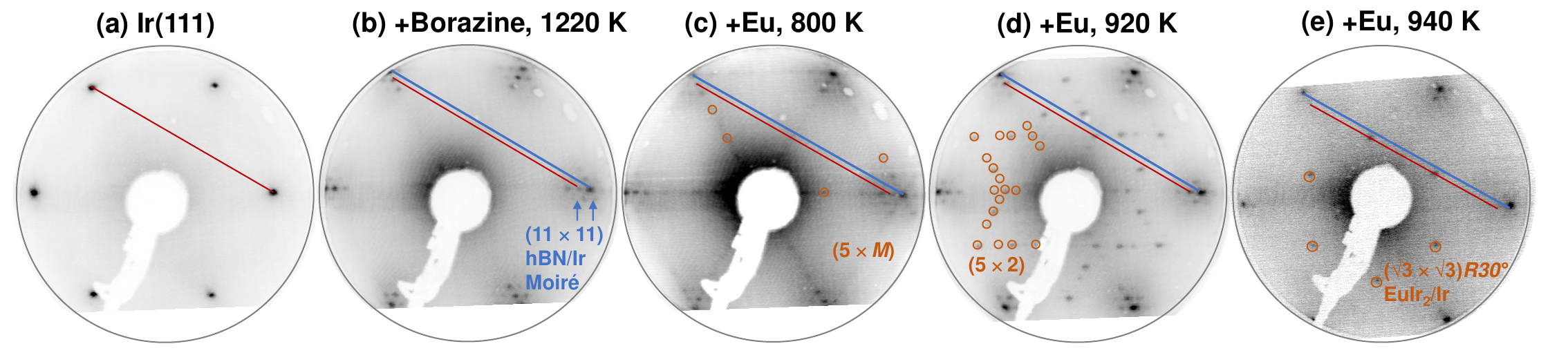}
    \caption{Low energy electron diffraction (LEED) analysis of the experiment. (a) clean Ir(111), (b) after exposure to borazine at $T_\textnormal{Sample}$ = 1220 K, a Moir{\'e} pattern appears, consisting of a (12 $\times$ 12) hBN unit cell on (11 $\times$ 11) Ir atoms. Eu intercalation was carried out at different temperatures on freshly prepared hBN/Ir(111): (c) at $T_\textnormal{Sample}$ = 800 K, a (5 $\times$ $M$) superstructure forms, where $M$ $>$ 2; (d) at $T_\textnormal{Sample}$ = 920 K, a (5 $\times$ 2) superstructure is observed; and (e) at $T_\textnormal{Sample}$ = 940 K, a ($\sqrt{3}$ $\times$ $\sqrt{3})R30^{\circ}$ superstructure is seen. LEED images were captured at a kinetic energy of 60 eV. The lines in part of the patterns serve as a guide to the eye to distinguish the spots.}
    \label{fig:2}
\end{figure*}

A thorough study utilizing LEED was performed both before and after the intercalation of Eu to examine the structural characteristics in detail. The LEED pattern of the pristine Ir(111) surface is displayed in Fig.~\ref{fig:2}(a), where a red line is used to connect two of the identical 3-fold Ir spots. Upon hBN growth at a temperature of 1220 K, a distinctive Moir{\'e} pattern emerges, characterized by a (12 $\times$ 12) hBN unit cell occupying (11 $\times$ 11) Ir atoms. This is consistent with the observations made by Farwick et al. \cite{farwick2016structure}, as highlighted in Fig.~\ref{fig:2}(b). The blue line connecting the most intense spots is characteristic of the hBN unit cell in reciprocal space. Next, we investigated the effects of different intercalation conditions after Eu deposition process through experiments conducted at various substrate temperatures, as depicted in Figs.~\ref{fig:2}(c)-(e). Temperature variations significantly alter the interface structure, leading to varying coverage. This observation aligns with findings by Bakhit et al. \cite{bakhit2023ferromagnetic} and Schumacher et al. \cite{schumacher2014europium}. They emphasize the substantial influence of thermal factors on the process of Eu intercalation.

At low Eu coverage of about 0.1 ML and substrate temperature 800 K, a (5 $\times$ $M$) superstructure was observed by LEED with respect to the Ir(111) substrate, see Fig.~\ref{fig:2}(c). A coincidence lattice appears exclusively along the $<$11$\bar{2}>$ directions; however, achieving uniform alignment along the $<\bar{1}$10$>$ directions is detained at low coverage, resulting in blurred LEED spots due to disorder. The parameter $M$, indicating periodicity along the $<\bar{1}$10$>$ directions, suggests a random distribution of Eu atoms, occupying spaces between every 3, 4, or 5 Ir atoms. This irregularity results in less defined hBN/Ir Moir{\'e} spots along the $<\bar{1}$10$>$ directions, although the overall Moir{\'e} pattern remains unchanged.
Further exposure of the (Eu)/hBN/Ir system to Eu at 920 K led to a compression of the intercalated layer, forming a (5 $\times$ 2) superstructure, as depicted in Fig.~\ref{fig:2}(d). In the supplemental material \cite{SM}, Fig. 1 presents the LEED patterns of the (5 $\times$ 2) superstructure with variation in the electron kinetic energy.
This compression is attributed to an increased presence of Eu atoms along the $<\bar{1}$10$>$ directions. The higher amount of Eu no longer allows random occupation of Ir rows but a positioning every two Ir rows, indicating a stable superstructure relative to the Ir substrate. After air exposure and subsequent annealing at 765 K, one can still appreciate the rests of the (5 $\times$ 2) superstructure.

In a different preparation, Eu was deposited onto hBN/Ir(111) holding the substrate at 940 K but this time slowly evaporating a critical Eu amount (30 min) to leave the Eu atoms just occupying surface positions and arrange in a ($\sqrt{3}$ $\times$ $\sqrt{3})R30^{\circ}$ superstructure as presented in Fig.~\ref{fig:2}(e). At this coverage, we observe the surface formation of EuIr$_{2}$ with a stoichiometry of (one Eu atom per two Ir atoms). Previous studies have demonstrated the appearance of such superstructures as an indication of the formation of rare earth-gold and rare earth-silver surface alloys with a similar one-to-two ratio \cite{ormaza2016high, que2020two, xu2020two, fernandez2020influence, ormaza2013laau, corso2010111, corso2010rare}. Moreover, the same superstructure was identified in a prior investigation of the Eu-Pt system \cite{bakhit2023ferromagnetic}. It has to be pointed out that the Eu spots are associated with the Ir substrate, not with the hBN overlayer that would produce superstructure spots on the blue line connecting hBN atoms. This situation is partially distinct from the Eu intercalation below graphene on Ir(111)~\cite{schumacher2014europium}. 

\begin{figure*}
    \centering
    \includegraphics[scale=0.6]{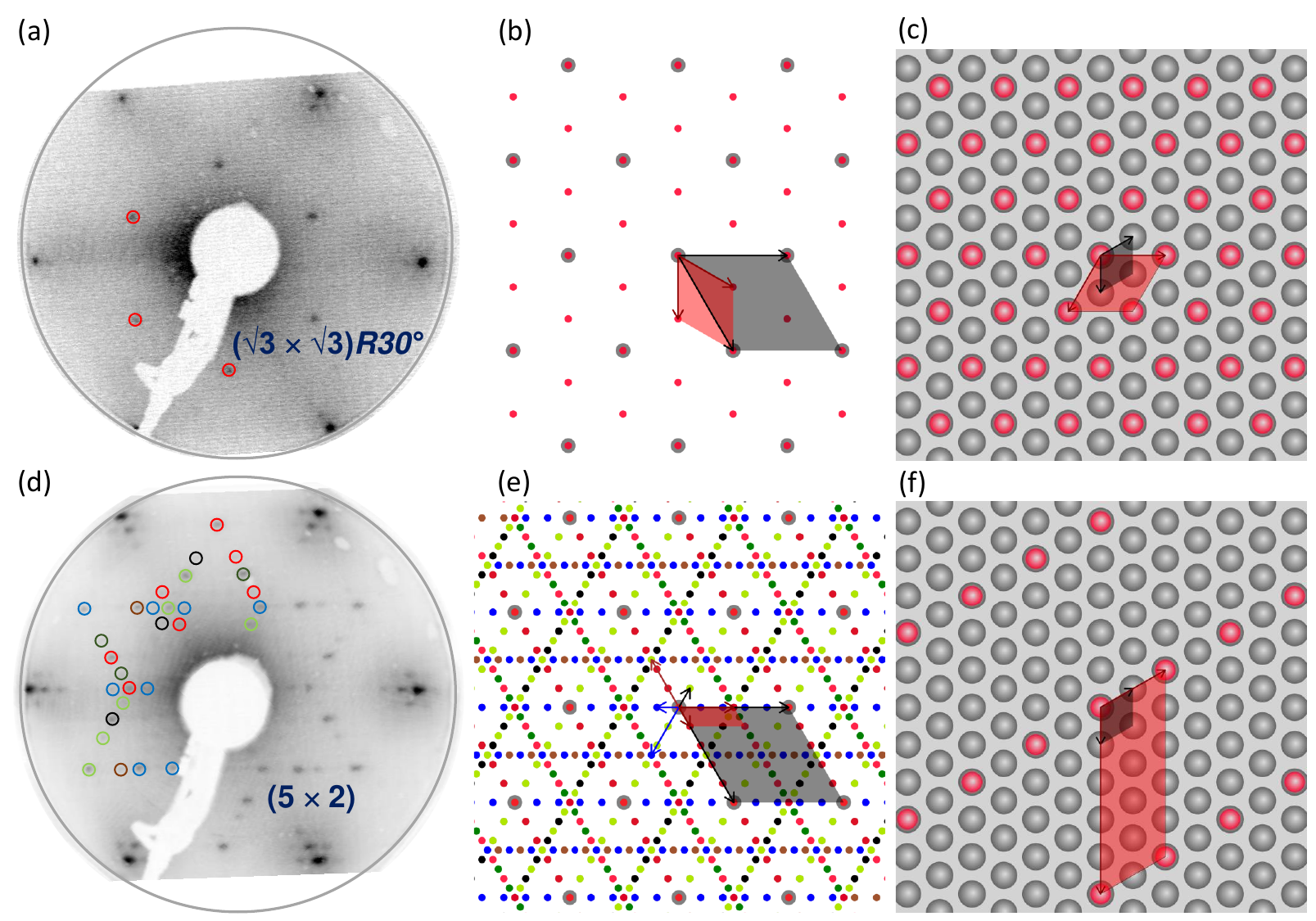}
    \caption{(a) LEED image taken after high-temperature Eu intercalation of 1/3 ML that reveals a ($\sqrt{3}$ $\times$ $\sqrt{3})R30^{\circ}$ superstructure, (b) the reciprocal lattice, and (c) the structural model in real space. For the lower temperature Eu intercalation, (d) presents the LEED image of the (5 $\times$ 2)  superstructure that corresponds to a superposition of three rotational domains (e) with one of the (5 $\times$ 2) reciprocal unit cells in red and the Ir lattice in grey. (f) Structural model of the (5 $\times$ 2) superstructure in real space. LEED images in (a) and (d) were taken at electron beam energies of 60 eV and 55 eV, respectively.}
    \label{fig:3}
\end{figure*}

The saturation coverage is one-third ML of Eu, leading to the formation of a $(\sqrt{3}~\times~\sqrt{3})R30^{\circ}$ superstructure with stoichiometry of one Eu atom per two Ir atoms displayed in Figs.~\ref{fig:3}(a) and (b) show its reciprocal lattice, while (c) presents the corresponding real-space structure. At 0.26 ML Eu coverage, a $(5~\times~2)$ superstructure appears, as seen in Fig.~\ref{fig:3}(d). This is evident in the STM analysis, further discussed in the text. The complexity of this superstructure is attributed to rational domains, illustrated in Fig. ~\ref{fig:3}(e) by the reciprocal lattice of the $(5~\times~2)$ (1 Eu atom per 9 Ir atoms) superstructure with three rotational domains at $120^{\circ}$ angle, alongside the Ir reciprocal lattice in black. The alignment between the theoretical model and the observed LEED pattern is obvious. Finally, Fig.~\ref{fig:3}(f) illustrates the real-space structure of the $(5~\times~2)$ superstructure, highlighting both the Ir and Eu unit cells.

The crystal structure of the EuIr$_{2}$ compound has been investigated, with the material found to adopt a Laves phase MgCu$_{2}$ type structure according to reports in the literature \cite{rp1965pr,pottgen1999syntheses}. The lattice parameter of this structure has been measured to be 7.565 nm \cite{tomuschat1984magnetic}, indicating a stable \cite{johnston1992structure} and well-defined crystallographic arrangement. Furthermore, the di-valent valence state of the Eu atoms within the EuIr$_{2}$ structure has been confirmed through previous studies \cite{bozorth1959magnetization}, highlighting the magnetic properties.
We believe that Eu has a di-valent character on the surface and tri-valent behaviour in the bulk. As has been found in the case of Eu on Ni studied by \cite{wieling2002electronic}.

\begin{figure*}
    \centering
    \includegraphics[scale=0.6]{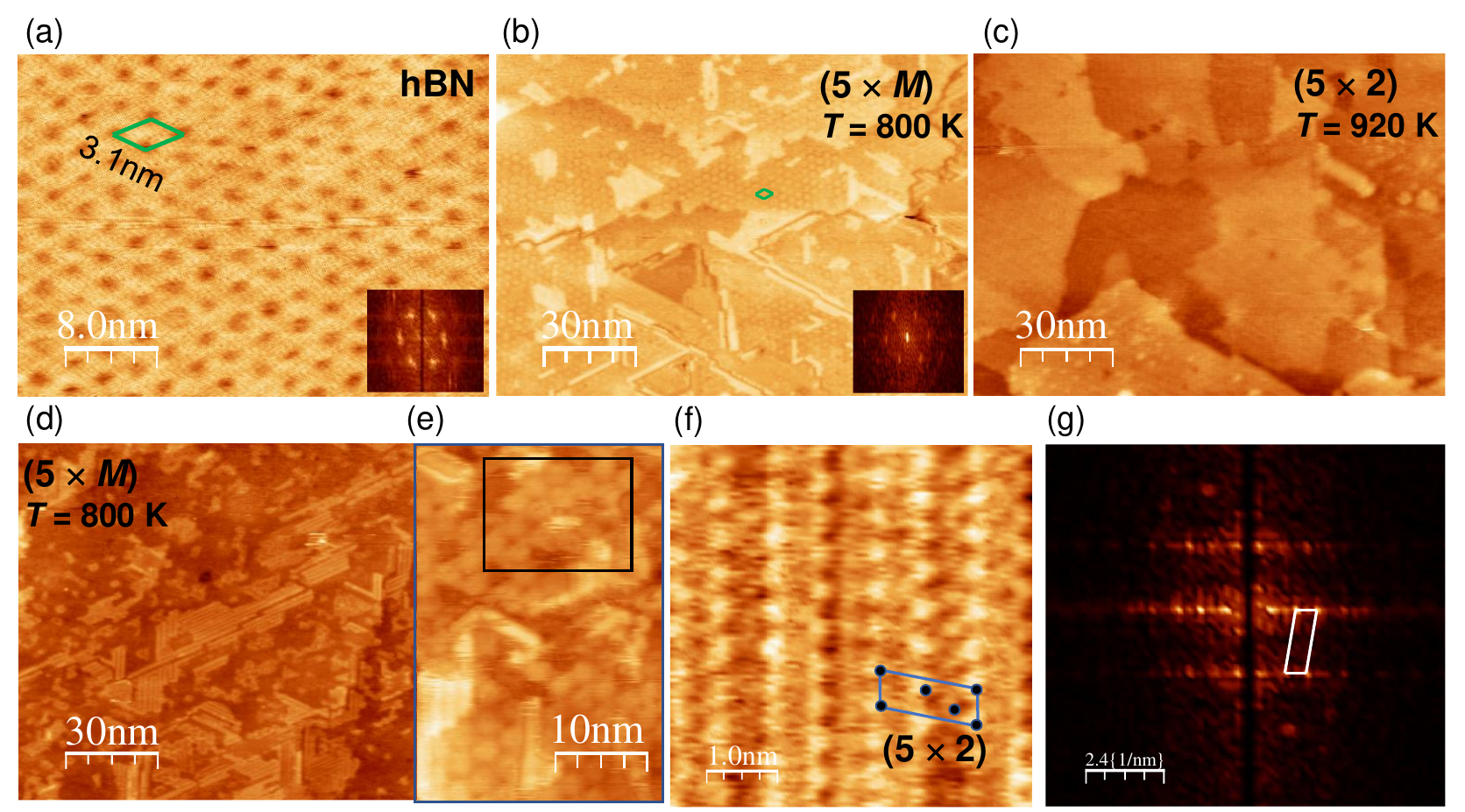}
    \caption{(a) STM image of hBN/Ir(111) Moir{\'e}, the green hexagon represents the Moir{\'e} lattice with lattice parameter (3.1 $\pm$ 0.1)nm and the inset shows an FT image containing a hexagonal pattern, corresponding to the hBN/Ir(111) Moir{\'e}. (b) After intercalation of 0.10 ML of Eu, the green hexagon shows the hBN/Ir(111) Moir{\'e} lattice, and the inset shows an FT image containing a hexagonal pattern, corresponding to the hBN/Ir(111) Moir{\'e}. (c) After depositing more than 1/3 ML of Eu. (d) and (e) After intercalation of 0.1ML of Eu each is taken at a different region; the highlighted area in black in (e) shows the hBN/Ir(111) Moir{\'e}. (f) Zoom in on one of the stripes in (d), showing the atomic resolution of $(5~\times~2)$ superstructure in blue together with a basis in green arrow. (g) FT image of (f). Tunneling parameters: $I$=0.13 nA; $U_{\text{bias}}$=0.5 V. }
    \label{fig:4}
\end{figure*}

STM images in Fig.~\ref{fig:4} provide detailed insights into the hBN/Ir(111) Moir{\'e} pattern and the effects of Eu intercalation on its topography. The distinctive Moir{\'e} unit cell, highlighted in green, is characterized by bright protrusions interspersed with slight central indentations. It has a lattice constant of (3.1 $\pm$ 0.1)nm, as confirmed by the Fourier-transformed (FT) image analysis in the inset of Fig.~\ref{fig:4}(a), further supporting the findings from Farwick et al. \cite{farwick2016structure}. However, a slight deviation from the expected theoretical Moir{\'e} periodicity of 3.25 nm indicates localized variations in the geometry of the hBN layer. These variations lead to the formation of wrinkles and defects that facilitate Eu intercalation, a phenomenon previously discussed in studies on Eu intercalation mechanisms beneath graphene on Ir(111) by Schumacher et al. \cite{schumacher2014europium}. The insights from these observations enhance our understanding of the structural properties and intercalation processes below 2D materials, a similar study on gold atom intercalation mechanisms by Daukiya et al. \cite{daukiya2019functionalization}.
Soon after the deposition of 0.10 ML of Eu, it passes through the wrinkles, forming patterns of stripes and islands, as seen in Fig.~\ref{fig:4}(b), where the brighter regions indicate intercalated areas. At low coverage, only 30$\%$ of the hBN layer is affected by Eu intercalation. Consequently, the hBN/Ir Moir{\'e} lattice remains observable, highlighted in green in Fig.~\ref{fig:4}(b), with the same Moir{\'e} lattice constant observed in panel (a). However, slight deviations are noted due to the strain induced on the hBN layer during intercalation \cite{schumacher2014europium}. The FT image is shown in the inset of Fig.~\ref{fig:4}(b).

Variations in the distribution of intercalated Eu result in uneven concentrations, particularly in areas where dense stripes are visible, as displayed in Fig.~\ref{fig:4}(d). The Moir{\'e} pattern varies slightly depending on the stripe density, as shown in Figs.~\ref{fig:4}(b)-(e). Upon closer inspection of one of these stripes in Fig.~\ref{fig:3}(d), a locally formed $(5~\times~2)$ superstructure was observed. This superstructure was undetected in the LEED pattern but is evident in Fig.~\ref{fig:4}(f), where the blue lattice represents the $(5~\times~2)$ unit cell, and the green indicates its basis. The FT of the $(5~\times~2)$ superstructure is shown in Fig.~\ref{fig:4}(g). However, a uniform distribution of Eu intercalation is achieved below hBN when increasing the Eu coverage up to 0.26 ML, resulting in Fig.~\ref{fig:4}(c).

\begin{figure*}
    \centering
    \includegraphics[scale=0.7]{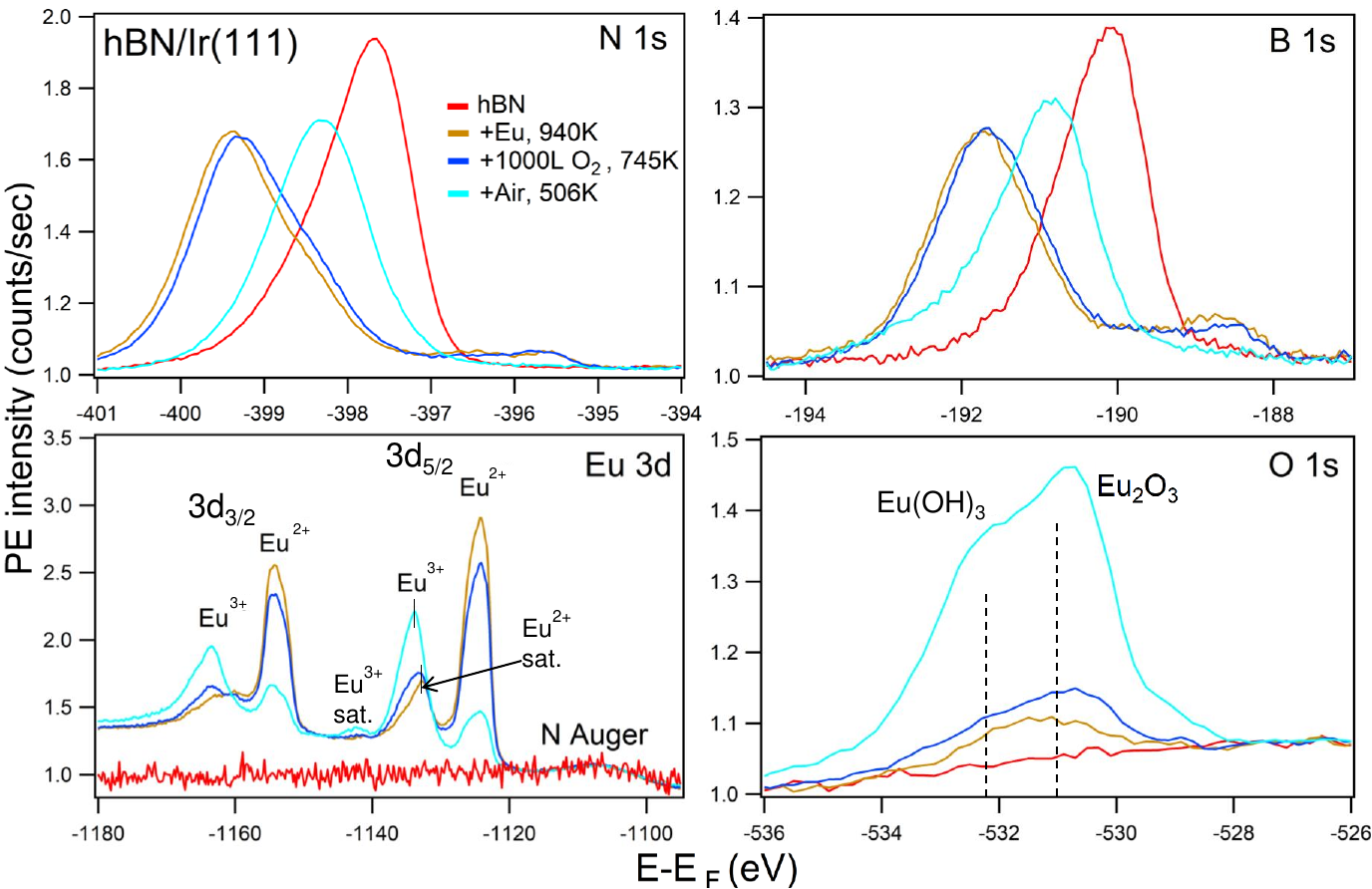}
    \caption{XPS analysis of Eu intercalation below hBN on the flat Ir(111) crystal. Displaying core level spectra of N 1s, B 1s, Eu 3d, and O 1s X-ray photoemission spectra taken at $h\nu$ = 1486.6 eV (Al K$\alpha$) for the 1/3 ML preparation before and after exposure to 1000 L of Oxygen, followed by exposure to ambient conditions.}
    \label{fig:5}
\end{figure*}

Core-level PE spectroscopy was utilized to examine the Eu intercalated layer beneath hBN on the Ir(111) substrate before and after intercalation, in order to understand its chemical composition and morphology. Investigating the impact of various Eu coverages on the chemical state of the system is of particular interest, as the Eu coverage is determined solely by the presence of di-valent Eu at the interface, as explained earlier.
At low coverage of 0.10 ML, the Eu 3d spectrum obtained, as illustrated in supplemental material \cite{SM}, top panel in Fig. 2, only displays a Eu$^{2+}$ configuration with a shake-up peak observed at a higher binding energy and no Eu$^{3+}$ contribution is detected. Furthermore, the O 1s spectrum reveals no additional components compared to the pre-intercalation spectra, suggesting no oxygen presence. 
As Eu coverage increases to 0.26 ML, the Eu 3d spectrum shows equal amounts of Eu$^{2+}$ and Eu$^{3+}$ contributions, while the O 1s spectrum consists of Eu$_{2}$O$_{3}$ and Eu-OH components. These spectra are shown in the supplemental material \cite{SM}, bottom panel of Fig. 2. 

With a further increase to one-third of the ML, the Eu 3d spectrum displayed a mixture of valence states, specifically the presence of di-valent Eu$^{2+}$ and tri-valent Eu$^{3+}$, shown in Fig. \ref{fig:5}. On the one hand, the existence of Eu$^{2+}$ suggests the development of a surface compound EuIr$_{2}$ has been confirmed by the LEED analysis. On the other hand, the Eu$^{3+}$ peak, which includes 18$\%$ shake-up of the Eu$^{2+}$\cite{laubschat1983final,schneider1981shake,cho1995origin}, remains prominent on the hBN layer. Before intercalation, the N 1s and B 1s spectra in Fig.~\ref{fig:5} exhibited a distinct asymmetric peak shape, showing two components \cite{preobrajenski2007monolayer} with a pronounced shoulder at higher binding energies \cite{farwick2016structure,orlando2012epitaxial,orlando2014epitaxial}. The substrate-overlayer interaction strength in this system falls in between the weak hBN on Pt(111) and the intermediate hBN on Rh(111) systems \cite{preobrajenski2007influence}. The same hBN quality was obtained in the second preparation acquired at a temperature of 1120 K.

Hence, in Fig.~\ref{fig:5}, the N 1s and B 1s spectra exhibit a noticeable broadening and a shift of around 1.7 eV and 1.65 eV, respectively, toward higher binding energies due to the formation of EuIr$_{2}$. Furthermore, a side peak detected at lower binding energies was interpreted to originate from single B and N atoms, consistent with the findings reported by~\cite{orlando2012epitaxial}. This was observed in both preparations. Upon exposure to 1000 L of O$_{2}$, a slight shift toward lower binding energies was detected in the N and B 1s spectra. However, these spectra retained the same characteristics as those inspected prior to exposure. The Eu 3d spectrum shows a notable decrease in Eu$^{2+}$ with a simultaneous increase in Eu$^{3+}$, indicating the formation of Eu$_{2}$O$_{3}$ and Eu-OH at binding energies of -532.3 eV and -531.0 eV, respectively. In addition, the EuO component at a binding energy of -529.6 eV was determined from the fitting results of the O 1s spectra after Eu intercalation, as shown in the supplemental material \cite{SM} in Fig. 4.
Subsequently, the system was exposed to ambient conditions followed by vacuum annealing at 506 K to remove some impurities. As a result, the N 1s and B 1s spectra began to recover their initial shape and position, indicating that the contact interface changed from a strong interaction with EuIr$_{2}$ to a weaker interaction with Eu$_{2}$O$_{3}$. Similar trends were noted in related systems, as reported in recent research \cite{bakhit2023ferromagnetic}. The overall intensity of the Eu 3d spectra in Fig.~\ref{fig:5} decreases significantly. It shows a doubling of the Eu$^{3+}$ amount, accompanied by a substantial reduction in the Eu$^{2+}$ amount. Moreover, the O 1s spectrum
shows a remarkable increase in both components Eu$_{2}$O$_{3}$ and Eu-OH and a decrease in the EuO component. The fit presented in supplemental material \cite{SM}, in Fig. 4, reveals a component with a binding energy of -533.0 eV corresponding to adsorbed water. Nonetheless, this accounts for the increase in the amount of Eu$^{3+}$, corresponding to the formation of Eu$_{2}$O$_{3}$.

\begin{figure*}
  \centering
   \includegraphics[scale=0.5]{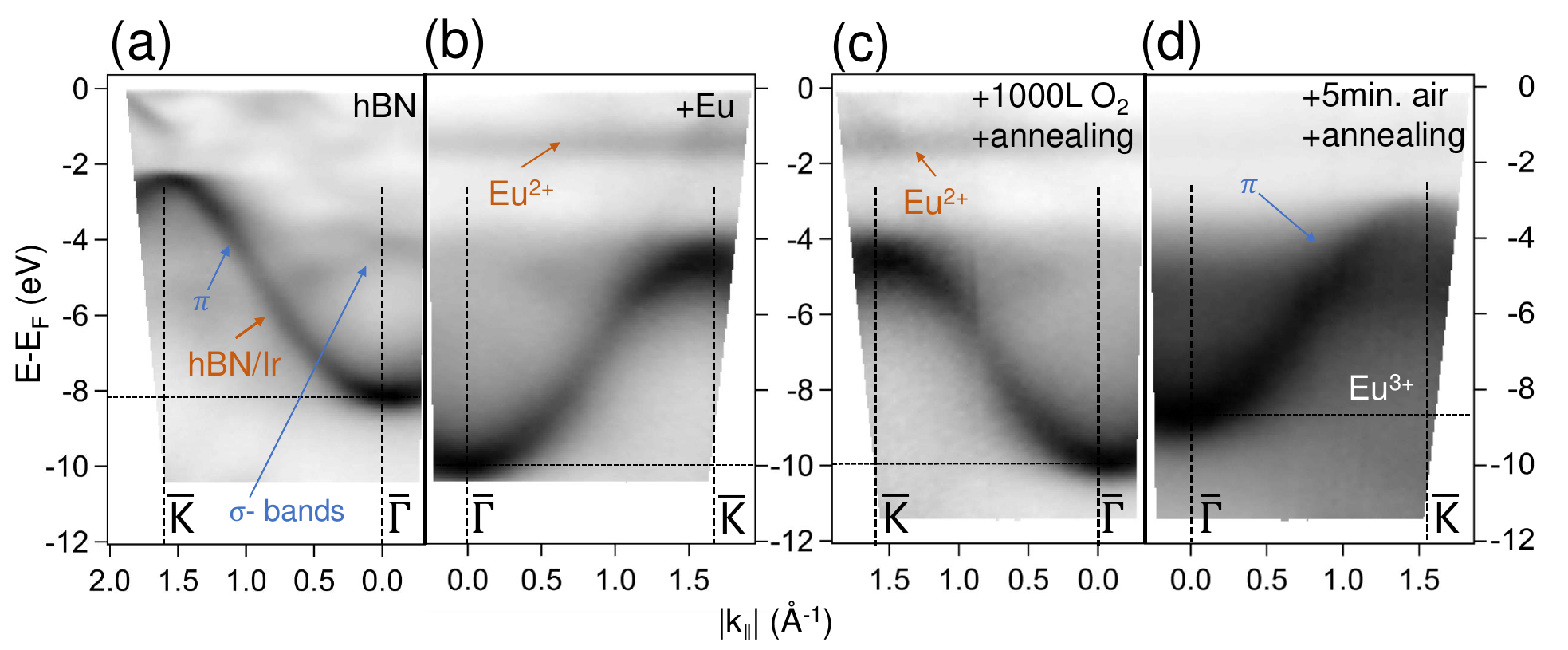}
    \caption{ARPES analysis of the band structure of the hBN layer prior and after Eu intercalation. He II $\alpha$ ($h\nu$ = 40.8 eV) photoemission intensity maps along [$\bar{\Gamma}$][$\bar{\mathrm{K}}$] direction of the hBN band structure for (a) hBN/Ir(111), (b) after intercalation of one-third ML of Eu at $T$ = 940 K, (c) after dosing 1000 L of O$_{2}$ and annealing $T$ = 745 K, and (d) followed by air exposure and annealing at $T$ = 506 K. The strongly dispersive band feature corresponds to the $\pi$-band of hBN at the indicated interfaces.}
    \label{fig:7}
\end{figure*}

Lastly, let us comprehensively address the electronic band structure of the (Eu)/hBN/Ir(111) system before and after air exposure. This was investigated using ARPES with He II$\alpha$ light at photon energy $h\nu$ = 40.8 eV, as shown in Fig.~\ref{fig:7}. This photon energy is very suitable for the hBN $\pi$ band emission but less suited for the $\sigma$ bands. The study particularly focused on the impact of air exposure on the hBN-protecting layer. In Fig.~\ref{fig:7}, the hBN $\pi$ band disperses from its minimum energy at $\bar{\Gamma}$ upwards towards the $\bar{\mathrm{K}}$ point where it reaches the closest energy to the Fermi level, representing the minimum and maximum energy of the surface Brillouin zone. This band extends from -8.0 to -2.4 eV upon hBN growth, while the Ir valence bands lie noticeably closer to the Fermi level. After intercalating one-third ML of Eu, the hBN $\pi$ band shifts by 2 eV to higher binding energies, and the di-valent Eu 4f emissions appear at approx. 1.6 eV. The $\pi$-band shift results due to the strong interaction of the hBN with the EuIr$_{2}$ interface; such $\pi$ band shifts were observed in other systems \cite{pervan2015li, jolie2014confinement,larciprete2012oxygen,schroder2016core}. Upon exposure to 1000 L of oxygen, the $\pi$ band of the hBN exhibited a slight shift, while the di-valent Eu 4f intensity is maintained. However, all bands became somewhat blurred after the air exposure. Despite this, the hBN $\pi$ band is still observed, with a notable shift toward lower binding energies, i.e., a partial recovery of the hBN $\pi$ band. The di-valent Eu 4f emission is nearly invisible, but the tri-valent Eu 4f emission appears between -4 and -10 eV. Our observations indicate that neither oxygen nor air exposure affects the hBN layer, which remains intact. This behavior looks similar to the (Eu)/hBN/Pt(111) system \cite{bakhit2023ferromagnetic}.

\section{Conflicts of interest}

There are no conflicts to declare.

\section{acknowledgments}

We acknowledge the financial support received from the Spanish MCIN/AEI/10.13039/501100011033 through grant PID2020-116093RB-C44, as well as the Basque Government through grant IT-1591-22 and Gipuzkoa Next program of the Diputaci{\'o}n Foral de Gipuzkoa DFG- 2023-CIEN-000077. Additionally, this research was partially funded by the CALIPSOplus project (Grant Agreement 730872) under the EU Horizon 2020 Framework Programme for Research and Innovation. We also express our gratitude to San Sebasti\'an Fomento and COST-action OPERA for their funding contributions.

\section{Summary}

We have examined the structural and electronic properties of Eu intercalated beneath hBN on an Ir(111) substrate at coverage less than one-third of a monolayer. Analysis using LEED indicated the presence of distinct superstructures, specifically (5 $\times$ $M$), (5 $\times$ 2), and $(\sqrt{3} \times \sqrt{3})R30$ depending on the coverage and preparation temperature. The (5 $\times$ M) superstructure, resulting from 0.1 ML of Eu coverage, exhibits a unidirectional ordering of Eu atoms due to the low coverage. It is a precursor of the higher coverage (5 $\times$ 2) phase with 3 Eu atoms in the unit cell. Finally, the $(\sqrt{3} \times \sqrt{3})R30$ corresponds to 1/3 ML Eu coverage in a EuIr$_2$ surface compound. In all cases, the Eu is below the hBN layer with the interface atoms in a di-valent configuration, but the Eu atoms that diffuse into the Ir bulk have a 3+ configuration. The hBN layer remains intact after intercalation, presenting an energy shift due to the stronger interaction of BN with the Eu atoms. Eu below the hBN layer is partially protected from oxidation, even after air exposure, quite in contrast to an unprotected Eu-Ir alloy that already presents degradation under UHV conditions.

\bibliography{main}

\end{document}


\title{Formation and protection of a Eu–Ir surface compound grown below hexagonal boron nitride\protect\\
-Supplemental Material-}

\author{Alaa Mohammed Idris Bakhit$^{1,2}$}
 \email{alaa.mohammed$@$ehu.eus}

 \author{Khadiza Ali$^{3,4}$}
 
 \author{Frederik Schiller$^{1,5}$}
 \email{frederikmichael.schiller$@$ehu.es}
 
 \affiliation{%
$^1$Centro de F\'{\i}sica de Materiales (CSIC-UPV-EHU) and Materials Physics Center (MPC), 20018 San Sebasti\'an, Spain
}%
\affiliation{%
$^2$Departamento de F\'{\i}sica Aplicada I, Universidad del Pa\'{\i}s Vasco UPV/EHU, 20018 San Sebasti\'an, Spain
}%
\affiliation{%
$^3$Chalmers University of Technology, Göteborg, Chalmersplatsen 4, 412 96 Göteborg, Sweden 
}%
\affiliation{%
$^4$Department of Physics, BITS Pilani, Hyderabad Campus, Hyderabad, 500078, Telangana, India
}%
\affiliation{%
$^5$Donostia International Physics Center, 20018 Donostia-San Sebasti\'an, Spain
}%
\maketitle

\begin{figure*}[htb!]
    \centering
    \includegraphics[scale=0.65]{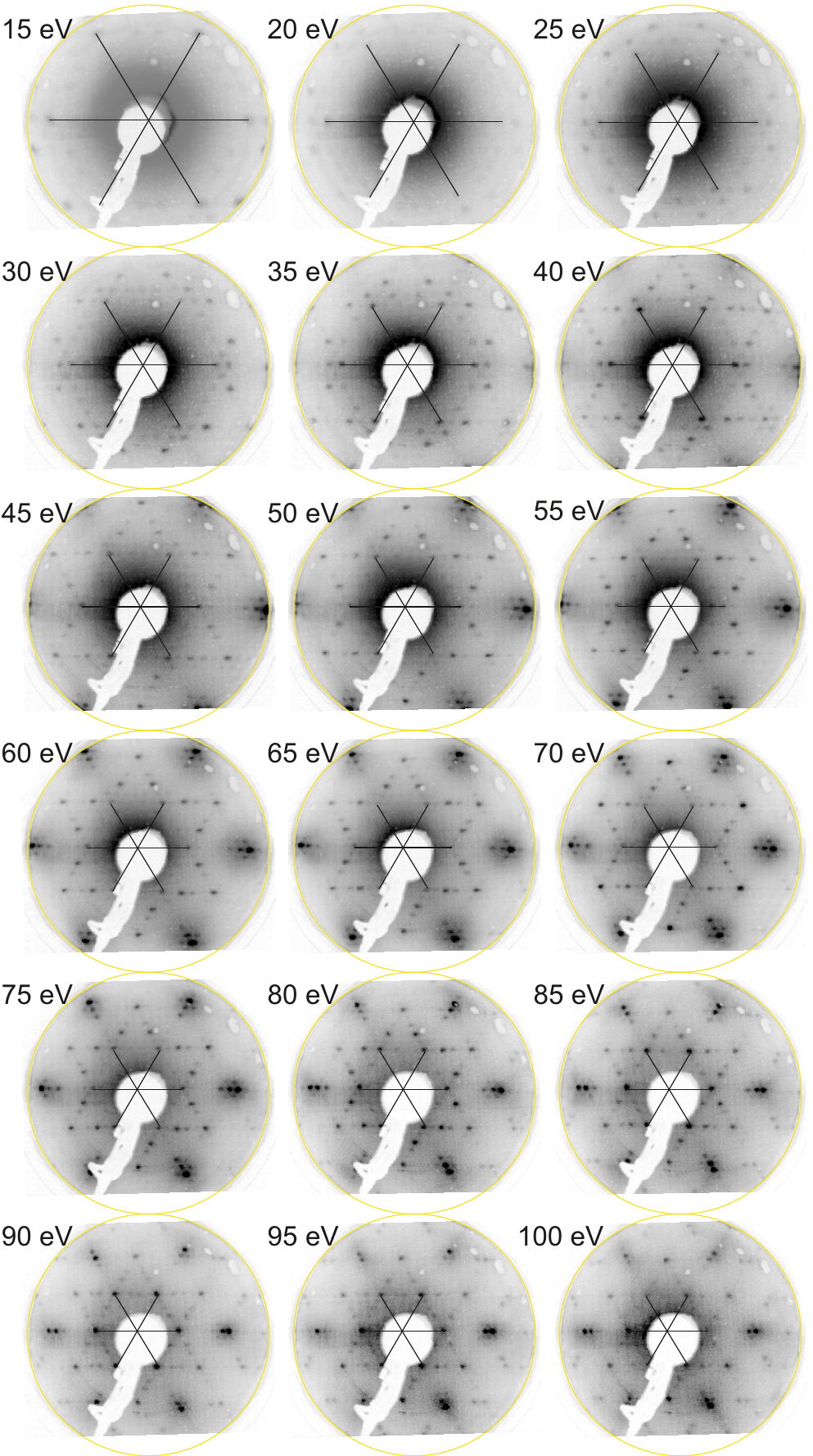}
    \caption{Low energy electron diffraction (LEED) of Eu intercalated below hBN on Ir(111) at T$_{\text{Sample}}$ = 920 K. A $(5 \times 2)$ superstructure is observed; an emergence of spots with energy variation. LEED images were captured at a kinetic energy range of 15 eV to 100 eV in steps of 5 eV. The lines correspond to a hypothetical $(2 \times 2)$ and serve as a guide for the eye to follow the evolution of the spots.}
    \label{fig:SI1}
\end{figure*}

Fig. \ref{fig:SI1} presents a systematic series of LEED patterns of the $(5 \times 2)$ superstructure of Eu intercalated below hBN at a substrate temperature of 920 K. The diffraction patterns were recorded over a range of incident electron kinetic energies, spanning from 15 to 100 eV in steps of 5 eV. These images capture the evolution of surface periodicity and diffraction spot intensity as a function of electron energy. The black lines are a guide to the eye for a hypothetical $(2 \times 2)$ structure. This set of LEED images provides evidence for the presence of a well-ordered $(5 \times 2)$ superstructure and illustrates the energy-dependent nature of the electron diffraction contrast.
\begin{figure*}[htb!]
    \centering
    \includegraphics[scale=0.65]{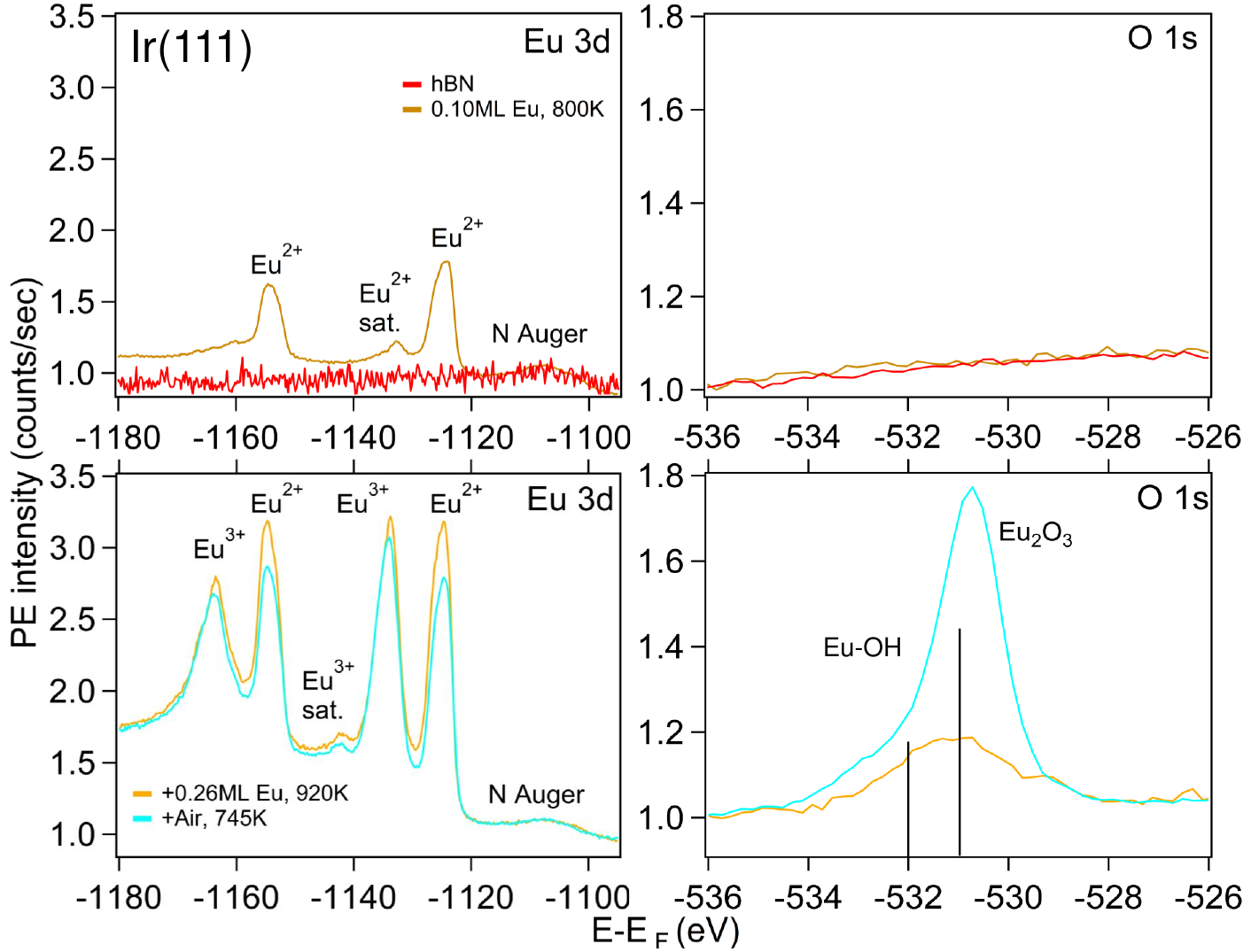}
    \caption{XPS spectra of Eu intercalation below hBN on the flat Ir(111) crystal. Core-level X-ray photoemission spectra of Eu 3d and O 1s taken at $h\nu$ = 1486.6 eV (Al K$\alpha$) for two Eu coverages are shown. Top: 0.10 ML Eu coverage prepared at substrate temperature of 800 K, and bottom: 0.26 ML Eu coverage at 920 K.}
    \label{fig:SI2}
\end{figure*}
At low Eu coverage of 0.10 ML, Eu 3d spectrum, as illustrated in Figure \ref{fig:SI2} top panel, resamples Eu in divalent contribution. This di-valent emission contains a satellite peak at E-E$_{\text{F}}$=-1132.8 eV, consistent with the observations reported by Cho et al. \cite{cho1995origin}. Furthermore, the O 1s spectrum reveals no components; thus, it is identical to the pre-intercalation spectrum and does not indicate oxygen contamination. Since Eu is very reactive even in ultra-high vacuum conditions, the lack of O 1s and tri-valent Eu indicated that all Eu went below the hBN.
As Eu coverage increases up to 0.26 ML, bottom panel of Figure \ref{fig:SI2}, the Eu 3d spectrum shows equal amounts of Eu$^{2+}$ and Eu$^{3+}$ contributions. The high amount of Eu$^{3+}$ is interpreted as due to diffusing Eu into the bulk forming EuIr$_{x}$ alloy ($x~>$~2). Due to the longer evaporation process, a small increase in O 1s is observed, compatible with the above-mentioned strong reactivity of Eu.
Similarly to the Pt case studied previously \cite{bakhit2023ferromagnetic}, the O 1s spectra exhibit two components, Eu$_{2}$O$_{3}$ and Eu-OH, at binding energies previously identified. After exposure to air, the overall intensity of the Eu 3d spectrum decreases, especially a reduction in Eu$^{2+}$ due to oxidation of the surface compound EuIr$_{2}$. 
The overall intensity decreased slightly due to the overgrowing oxygen atoms. Note that the increase in O 1s is strong, compared to the small change in Eu 3d spectrum. This means that the Eu$^{3+}$ component prior to air exposure raised mainly from the bulk, not from the small oxidation species of Eu-OH or Eu$_{2}$O$_{3}$.
For this sample, after air exposure and soft annealing, in the O 1s spectrum, a predominant Eu$_{2}$O$_{3}$ component and a minor amount of Eu-OH were detected, confirming our interpretations. 
\begin{figure*}[htb!]
    \centering
    \includegraphics[scale=0.7]{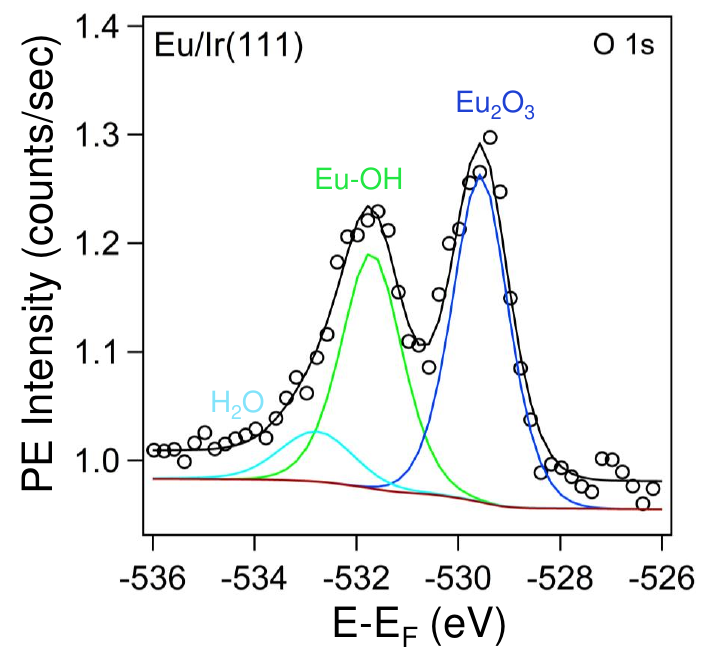}
    \caption{XPS peak fit analysis of Eu deposited on the flat Ir(111) crystal. The core-level spectrum of O 1s was taken at h$\nu$ = 1486.6 eV (Al K$\alpha$).}
    \label{fig:SI4}
\end{figure*}
Figure \ref{fig:SI4} shows the O 1s photoemission spectrum for Eu deposited on an Ir(111) surface acquired after 1 hour from the Eu deposition (see main text). The experimental data, represented by open circles, have been fitted using multiple Doniach–Šunjić (DS) line shapes \cite{SDoniach1970}, which account for the asymmetric line profiles typically observed in metallic systems. The overall fit, represented by a solid black line, comprises several components (coloured curves) that indicate the presence of chemically distinct oxygen species or environments. The spectrum reveals at least two prominent peaks at low and high binding energies corresponding to Eu$_{2}$O$_{3}$ and Eu-OH, respectively. The third component in the O 1s spectrum corresponds to adsorbate water with E-E$_{\text{F}}$= -533.0 eV.

\begin{figure*}[htb!]
    \centering
    \includegraphics[scale=0.7]{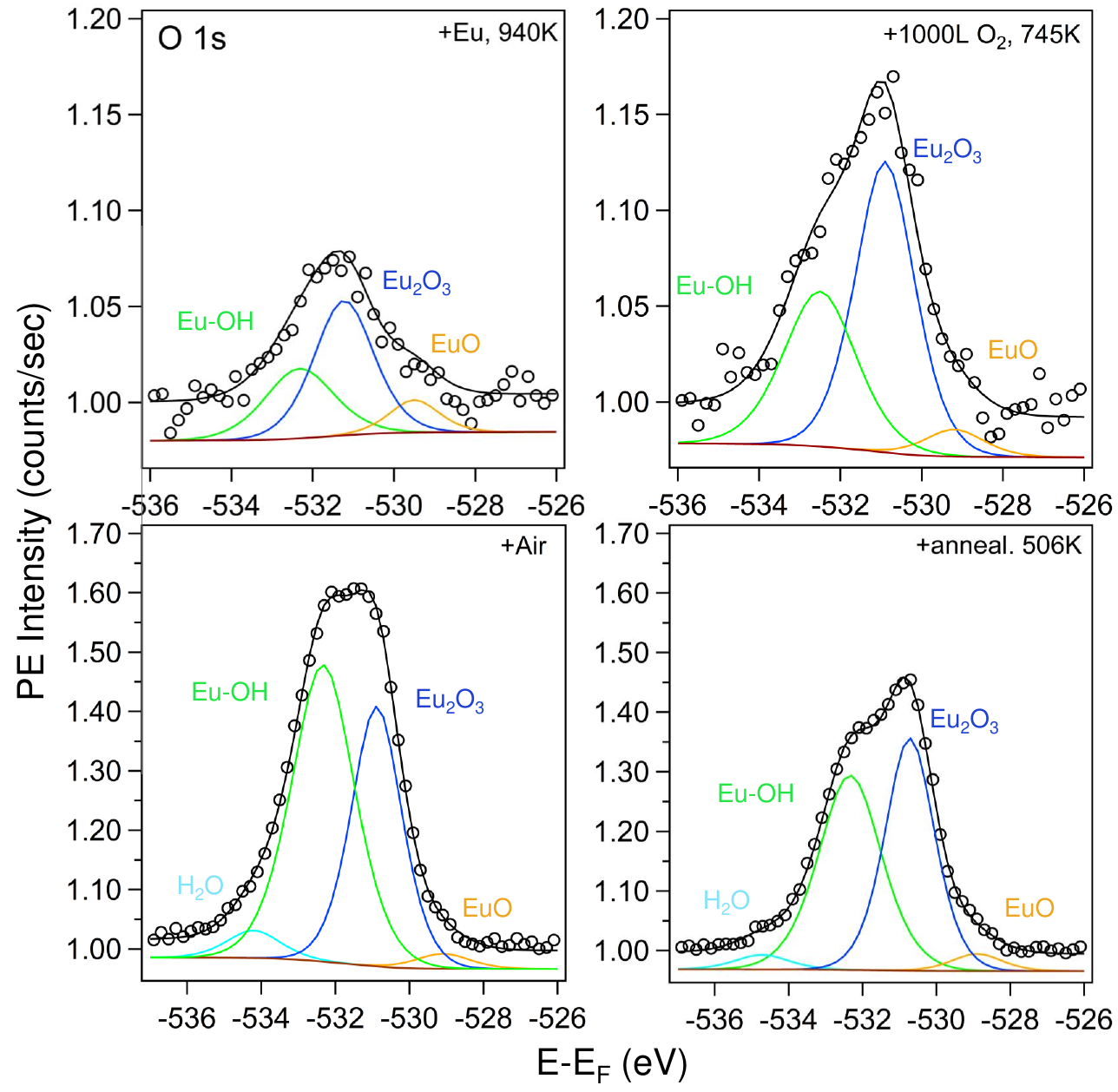}
    \caption{XPS peak fit analysis of Eu intercalation below hBN on the flat Ir(111) crystal. Displaying core-level spectra of O 1s were taken at h$\nu$ = 1486.6 eV (Al K$\alpha$) for the one-third ML preparation.}
    \label{fig:SI3}
\end{figure*}
The top panel of Figure \ref{fig:SI3} displays the O 1s core-level spectra obtained after intercalating one-third of a monolayer (ML) of Eu beneath hBN on Ir(111), followed by exposure to 1000 L of O$_{2}$, and annealing at 745 K. The spectrum reveals three distinct components corresponding to Eu-OH, Eu$_{2}$O$_{3}$, and EuO (with di-valent Eu) with energies of -532.3 eV, -531.0 eV, and -529.6 eV, respectively, to the Fermi level. 
After exposure to 1000 L of O$_{2}$, an overall increase in spectral intensity is observed, particularly for the Eu-OH, Eu$_{2}$O$_{3}$ components, while the EuO signal diminishes, indicating further oxidation of Eu species. The bottom panel shows the spectrum after air exposure, where a significant rise in O 1s intensity occurs. This spectrum now includes four components, attributed to adsorbed water, Eu-OH, Eu$_{2}$O$_{3}$, and EuO. Subsequent vacuum annealing up to 506 K reduces the overall intensity of the O 1s peak, suggesting partial desorption of surface contaminants introduced during air exposure.

Lastly, we want to mention an overall peak shift of approx. 1eV between the oxidized Eu/Ir system and the oxidized hBN/Eu/Ir system towards higher binding energies. We interpret these shifts as due to a change in the work function of both systems and the fact that the oxygen core level is bound to the vacuum level. Note that Ir(111) has a work function of 5.74eV~\cite{Michaelson1977_JAP} and hBN/Ir reduces this value to $\approx$4.2-4.4eV~\cite{Schulz2014_PRB}.

\bibliography{main}